\documentclass[journal,comsoc]{IEEEtran}
\usepackage[T1]{fontenc}
\usepackage{amsmath}
\usepackage[cmintegrals]{newtxmath}
\usepackage{url}
\usepackage{soul,color,graphicx,float,epstopdf}
\usepackage{tablefootnote,textcomp,gensymb}
\usepackage{eurosym,booktabs,multirow}
\usepackage[caption=false]{subfig}
\usepackage{amsfonts} 
\usepackage{algorithmic}
\usepackage{array}
\usepackage{stfloats}
\usepackage{mathtools}
\usepackage{pdfpages}
\usepackage{balance}
\usepackage{xcolor}
\usepackage{comment,enumitem}
\usepackage[normalem]{ulem}
\usepackage[implicit=false]{hyperref}

\definecolor{Orange}{rgb}{1,0.5,0}
\newcommand{\todo}[1]{\textsf{\textbf{\textcolor{Orange}{[#1]}}}}

\begin{document}
\title{Large Generative AI Models meet Open Networks for 6G: Integration, Platform, and Monetization}

\author{
Peizheng~Li,~\IEEEmembership{}
         Adrián Sánchez-Mompó,~\IEEEmembership{}
         Tim Farnham,~\IEEEmembership{}
         Aftab~Khan,~\IEEEmembership{}
         Adnan~Aijaz~\IEEEmembership{}
 \thanks{The authors are with the Bristol Research and Innovation Laboratory, Toshiba Europe Ltd., U.K. (e-mail: peizheng.li@toshiba-bril.com).

This work was supported partially by the 6G-GOALS project under the 6G SNS-JU
Horizon program, n.101139232.
}}

\maketitle

\begin{abstract}
Generative artificial intelligence (GAI) has emerged as a pivotal technology for content generation, reasoning, and decision-making, making it a promising solution on the 6G stage characterized by openness, connected intelligence, and service democratization.
This article explores strategies for integrating and monetizing GAI within future open 6G networks, mainly from the perspectives of mobile network operators (MNOs). We propose a novel API-centric telecoms GAI marketplace platform, designed to serve as a central hub for deploying, managing, and monetizing diverse GAI services directly within the network. This platform underpins a flexible and interoperable ecosystem, enhances service delivery, and facilitates seamless integration of GAI capabilities across various network segments, thereby enabling new revenue streams through customer-centric generative services. Results from experimental evaluation in an end-to-end Open RAN testbed, show the latency benefits of this platform for local large language model (LLM) deployment, by comparing token timing for various generated lengths with cloud-based general-purpose LLMs. Lastly, the article discusses key considerations for implementing the GAI marketplace within 6G networks, including monetization strategy, regulatory, management, and service platform aspects.
\end{abstract}


\section{Introduction}
\label{sec:intro}
Generative artificial intelligence (GAI) has emerged as a compelling and prominent research area due to its proven success in content generation services. Large GAI models, such as large language models (LLMs), image and video generation models, and multi-modality models, excel at understanding language and performing general-purpose tasks.

Models like GPT-4, Deepseek, Gemini, LLaMA, and Claude demonstrate powerful capabilities in context understanding, planning, responding, and code generation. These models can be customized for specific industries using techniques like retrieval-augmented generation (RAG), low-rank adaptation (LoRA), and prompt-tuning for cost-effective updates.

\textcolor{black}{Integration of AI and communication networks is at the heart of 6G evolution, as highlighted in the “IMT-2030 Framework''~\cite{ituimt2030}.}
GAI, especially through LLMs~\cite{bariah2024large}, is seen as a \textcolor{black}{key} enabler of this integration, enhancing wireless communication systems with advanced understanding, reasoning, and generating capabilities -- essential for developing in-network intelligence.

GAI is increasingly being adopted as a service in the telecoms sector. For example, LLMs are used for analyzing 3GPP specifications, data synthesis, anomaly detection, network modelling, and personalized service generation. Image generation models are applied in the goal-oriented and semantic communication paradigm for image reconstruction.

GAI services are expected to significantly impact the socioeconomics of the telecoms industry~\cite{maatouk2024large}, by delivering customer-centric services, improving operational efficiency, creating new products and revenue streams, reducing costs, and fostering innovation. It should be noted that the role and practices of mobile network operators (MNOs) are crucial in this evolution.

However, integrating GAI with telecoms \textcolor{black}{systems and} networks remains challenging for MNOs. 
Ongoing GAI-focused research is driven by computer sciences rather than telecoms and is generally confined to academic and AI circles. While MNOs can access GAI capabilities for potential services like conversational chatbots~\cite{maatouk2024large} via cloud platforms (e.g., Google's), such solutions are not attractive due to limited prospects of intrinsic control and customization of GAI services. Three key issues need to be addressed in this respect.

\begin{itemize}
    \item \emph{Lack of fine-grained control.} Third-party cloud solutions often function as black boxes, limiting an MNO's ability to enforce domain-specific policies, guarantee low-latency responses, or optimize model usage. 
    \item \emph{Scalability and cost structure.} Large GAI models can be very costly to train and deploy at scale. Offloading tasks to generic cloud LLMs may not produce the desired return on investment for telecom-specific usage. 
    \item \emph{Fragmented business models.} As GAI moves in-network, new revenue streams arise (e.g., user-facing AR/VR experiences, generative chatbots for enterprise customers), necessitating robust billing and service orchestration frameworks.
\end{itemize}

\textcolor{black}{The key principles of open radio access network (Open RAN), i.e., openness, disaggregation, cloudification, and programmability, hold transformative potential for 6G architectural evolution.}
The traditional monolithic RAN is \textcolor{black}{embracing} disaggregated RAN components~\cite{10329947} while MNOs are increasingly adopting cloud-neutral platforms supporting multi-cloud, private cloud, and hybrid configurations. \textcolor{black}{While} the shift towards openness complicates network management, especially with AI integration across \textcolor{black}{RAN, edge, and cloud layers, it also provides opportunities for innovation.}

As AI evolves into GAI and 6G networks become more open, there is a growing need to address these issues and explore monetization strategies for GAI within open networks. This article \textcolor{black}{investigates} a marketplace strategy for integrating GAI in networks, focusing on collaboration, service access and management, monetization, and the development of GAI-native 6G open networks. \textcolor{black}{To our knowledge, this is one of the first studies proposing a monetization strategy for GAI models integrated into 6G networks, with telecoms GAI marketplace in the spotlight.}

The main contributions of this article are summarized as follows:
\begin{itemize}
    \item We design and implement an API-centric telecoms GAI marketplace platform, serving as the entry point for heterogeneous GAI services deployed across various network segments and the exit for integrated and meshed GAI services. This marks a significant step toward a (G)AI-native network, enabling seamless AI integration within telecom infrastructure.
    \item We demonstrate an in-network GAI deployment use case within an end-to-end Open RAN network, wherein an LLMs-based generative service is examined, highlighting the benefits of this approach in terms of reduced service latency compared to general-purpose cloud-based GAI services.
    \item We provide a detailed discussion on the marketplace framework, covering aspects such as service access, monetizing, regulation, management, and open service platform.
\end{itemize}

\section{Background and challeneges}
\subsection{Open Networks} 
In 6G context, an open network is characterized by an architecture that promotes interoperability and flexibility by adhering to open standards and interfaces across the RAN, core network, and cloud infrastructure. This approach facilitates the seamless integration of components from various vendors, supports modular and programmable network architectures, and fosters a competitive ecosystem by avoiding vendor lock-in.
Open networks include the following features:
\begin{itemize}
    \item Vendor-Neutral Infrastructure: Supports interoperability between equipment and software from multiple vendors, allowing MNOs to build diversified networks. 
    \item Multi-Cloud Flexibility: Enables the network to operate seamlessly across multiple cloud service providers, including public, private, and hybrid environments. This enhances network resilience, optimizes performance through strategic workload distribution, and offers cost management benefits through varied pricing models.
    \item Third-Party App Deployment: Allows external developers to deploy applications on the network, diversifying the application ecosystem. This enhances customization, meets specific needs, and opens new revenue opportunities through third-party partnerships.
\end{itemize}

\subsection{Overview of the SOTA Large Models}
\label{subsec:sota_agi_overview}
Large GAI models are advanced neural networks with billions of parameters, trained on vast datasets using powerful computational resources. They generate human-like data (text, images, videos) using transformer and attention mechanisms. These state-of-the-art models are actively researched and applied in language, image/video, and multi-modality generation.

Early milestones in GAI include OpenAI's GPT-3, which significantly advanced natural language processing and generation, showcasing the potential of LLMs across various applications. Building on this foundation, GPT-4 further enhanced the field by improving the model's ability to understand and generate human-like text.

In image generation, a major breakthrough was achieved with DALL-E, which introduced the capability to generate images from textual descriptions. The subsequent release of DALL-E 3 excelled in creating detailed and coherent images, making it an invaluable tool for artists and designers. Stable Diffusion further refined text-to-image generation by producing high-quality, detailed images from textual inputs. 
Other notable GAI services available in the market include Deepseek, Claude, Gemini, and LLaMA.

These generative models are generally designed for general-purpose use, targeting a wide range of users and applicable to various tasks across different fields. Meanwhile, purpose-specific GAI models are emerging, trained or fine-tuned with specific knowledge in vertical domains. For instance, Codex powers GitHub Copilot, providing developers with intelligent code completion and suggestions. In the communication and network domain, NetGPT~\cite{meng2023netgpt} has been developed for understanding and generating network traffic; Autonomous edge AI is raised in~\cite{shen2024large} for connected intelligence by using LLMs.

Integrating GAI models with Open Networks offers significant benefits for both UEs and MNOs. Near real-time generative services can be delivered to UEs directly from the network, enabling low-latency applications such as conversational AI, augmented reality, and immersive media. Meanwhile, the network’s operation and management, such as resource allocation, network slicing, and reconfiguration, can be further automated and optimized through the reasoning and decision-making capabilities of in-network GAI models, improving efficiency and adaptability.

\subsection{Cloud-based GAI Pricing Schemes}

Cloud-based GAI usually charges per API call, token count, or monthly subscription. Additional fees often apply for advanced fine-tuning or domain adaptation. However, standard billing mechanisms do not necessarily map well to in-network services where MNOs might need:
\begin{itemize}
    \item \emph{End-to-end resource accounting} (e.g., GPU usage at edge nodes and RANs, memory, bandwidth). 
    \item \emph{Bundled telecom services} (e.g., 5G/6G data packages plus generative chatbot subscriptions). 
    \item \emph{Regulatory compliance} around data locality and real-time performance guarantees.
\end{itemize}

\subsection{Marketplace}

Marketplaces connect buyers and sellers in a digital platform, facilitating the discovery, purchase, and delivery of products or services, such as popular cloud marketplaces (e.g., AWS, Azure) and app stores (e.g., Apple, Google). In a telecom GAI marketplace, participants may include:
\begin{itemize}
    \item MNOs providing network infrastructure and hosting capabilities. 
    \item Third-party GAI developers publishing models for specialized tasks. 
    \item Enterprise, xApps or end-users purchasing generative services as a subscription or pay-per-use model. 
\end{itemize}
Unlike general-purpose marketplaces, a telecom-focused GAI marketplace must handle tight service level agreements (SLAs), real-time traffic, and multi-tenant resource management.

\begin{figure*}[t]   
    \subfloat[\label{fig: marketplace integration}]{
      \begin{minipage}[t]{0.95\linewidth}
        \centering 
        \includegraphics[width=6.5in]{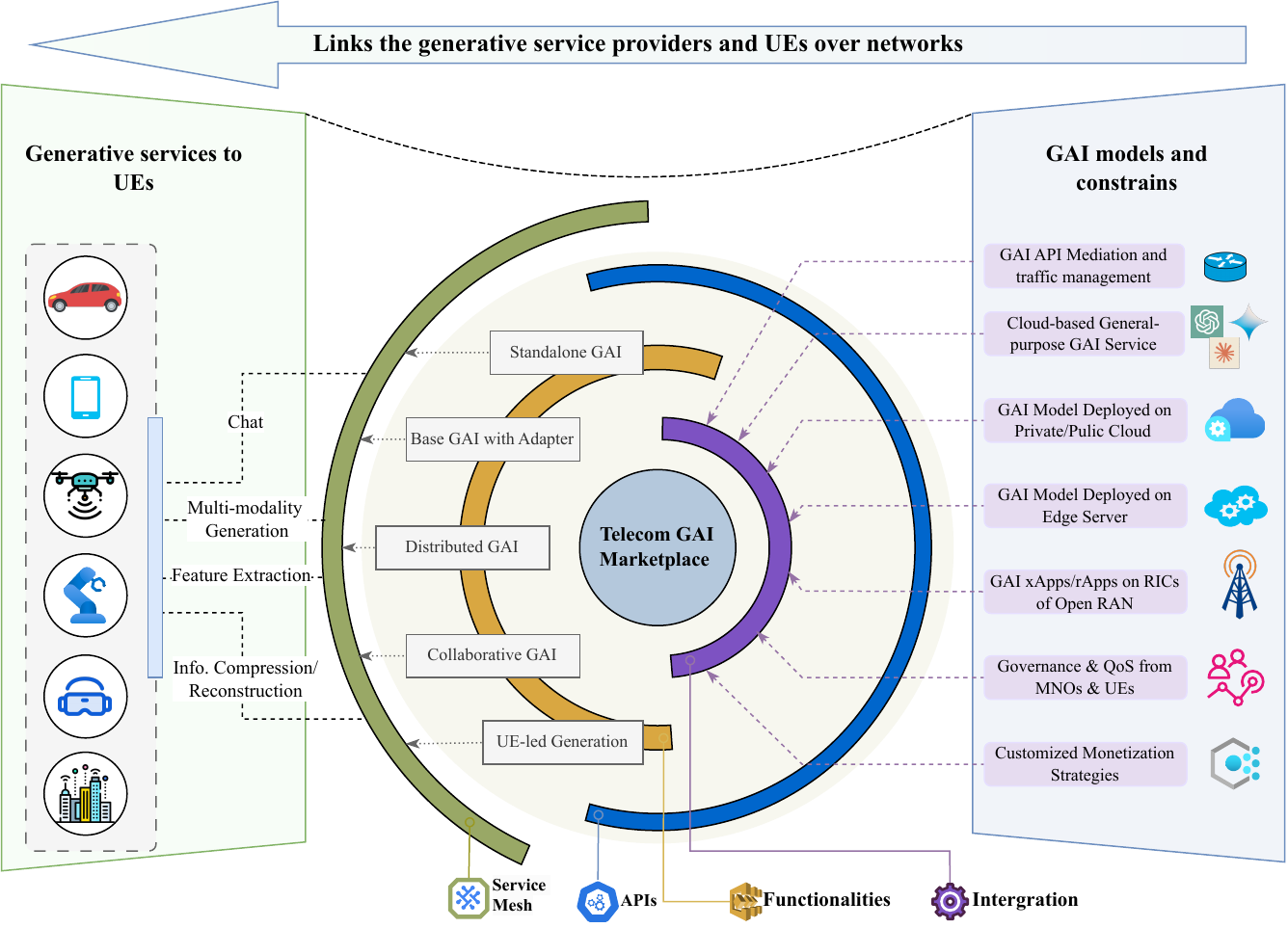}   
      \end{minipage}%
      }
      \\
        \subfloat[\label{fig: marketplace monetization process}]{
      \begin{minipage}[t]{0.95\linewidth}   
        \centering   
        \includegraphics[width=6.5in]{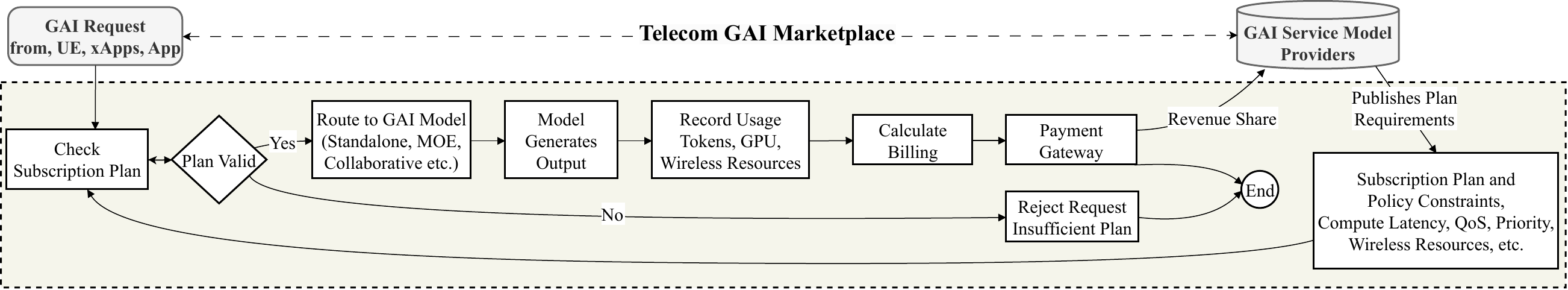}  
      \end{minipage} 
      }
      \caption{(a) Illustration of our conceptual telecom marketplace platform for GAI model integration and service delivery. It supports model registration, container-based deployment, multi-cloud integration, and open APIs for usage and billing; (b) Flowchart illustrating the monetization process in the telecom GAI marketplace)
      } \label{fig:marketplace_illustration}
\vspace{-1.45mm}
\end{figure*}

\subsection{Key Challenges from the MNO Perspective}
\label{sec:challenges}
Integrating GAI with open networks demands holistic solutions, especially from an MNO’s viewpoint:

\subsubsection{Monetization and Return on Investment (ROI)}
Deploying large GAI is expensive and time-consuming. MNOs need robust \emph{revenue models} that leverage AI-driven services to justify capital and operational expenditures. For example, they may bundle GAI capabilities into premium data plans or enterprise solutions.

\subsubsection{Control and Management Capability}
Cloud-based solutions often offer limited control or transparency. MNOs require fine-grained oversight of GAI services to ensure compliance with telecom-grade reliability, security, and QoS requirements. They need also to orchestrate these services dynamically across RAN, edge, and core layers.

\subsubsection{Service Quality and Latency}
GAI models can degrade end-user experience if requests are routed to distant cloud regions. Low latency, essential for interactive or real-time AI tasks, pushes GAI to be deployed closer to the user—i.e., at the network edge. MNOs thus need flexible, multi-cloud or edge-aware GAI orchestration to ensure predictable performance.

\subsubsection{Regulatory and Data Privacy}
Telecom data may contain sensitive user information. Guaranteeing compliance with privacy regulations (e.g., GDPR) and local data sovereignty requirements is crucial. MNOs must manage how and where GAI inferences occur and how user data is anonymized or protected in the pipeline.

Addressing these challenges calls for an approach that supports diverse GAI models with explicit monetization, strong management primitives, and policy compliance. The next sections detail a marketplace-based solution aligning with open network principles.

\section{Potential Methods of GAI-Open Network Integration}

Before introducing the marketplace, we briefly review prominent strategies for integrating large GAI models into telecom networks. These strategies hint at the breadth of design considerations (hardware, deployment, data) required for successful GAI operations in 6G.

\subsection{AI-native Network Architecture}
AI-native network architecture design is a vision for integrating GAI into 6G networks, which involves embedding AI capabilities into network stacks, ensuring that AI-driven functionalities are an integral part of network operations.  
For instance, one study~\cite{yang20226g} explores an architecture that coordinates cloud AI, edge AI, and network AI to deliver intelligent, customized services in future 6G networks. Another work~\cite{zhao2024open} introduces an open-source edge AI framework, embedding a AI plane within a multi-access edge computing framework. 

In this AI-native design, GAI can thus be launched or scaled automatically, though this approach alone might not resolve ecosystem or monetization hurdles.

\subsection{Goal-oriented and Semantic Communication}
Goal-oriented and semantic communication is an emerging paradigm that prioritizes the transmission of meaning over raw data~\cite{li2023open}, enabling more efficient and context-aware communication that surpasses traditional Shannon capacity. 
GAI is increasingly being considered as a backbone for semantic communication, serving as the semantic encoder/decoder and even as the knowledge base itself~\cite{xia2025generative}.

Although valuable for next-generation wireless efficiency, existing efforts of semantic communication often focus on local model integration rather than the broader scope of multi-stakeholder marketplaces.

\subsection{Marketplace-Based Integration}
\label{subsec:marketplace_integration}
To fully leverage open networks and third-party contributions, we propose a \emph{marketplace-based} approach. By using an API-driven platform, MNOs can unify management, integration, and monetization for a wide spectrum of GAI services -- ranging from stand-alone LLMs to collaborative inference pipelines. This strategy inherently supports multi-tenancy, flexible billing, resource orchestration, and regulatory compliance.

\section{Marketplace Solution for GAI and Open Network Integration}
We now present the conceptual architecture and implementation details of our \emph{telecoms GAI marketplace platform}. As illustrated in Fig.~\ref{fig: marketplace integration}, it offers a robust ecosystem wherein MNOs, third-party developers, and enterprise customers can seamlessly deploy and consume GAI services across disaggregated networks.

\subsection{Telecoms Marketplace Design Principles}
\begin{itemize}[leftmargin=*]
    \item \textbf{Standalone Model Support:} The marketplace manages and deploys large GAI models (e.g., LLMs or image generators) as self-contained services. Automated CI/CD pipelines allow MNOs or developers to release updates rapidly. 
    
    \item \textbf{Base Model with Adapter Support:} 
    To enable cost-efficient adaptation, the marketplace supports \emph{parameter-efficient fine-tuning} (PEFT) techniques such as LoRA, which attach small adapters to a general LLM. This helps domain specialists (e.g., enterprise developers) quickly customize GAI for telecom tasks, without full model retraining.

    \item \textbf{Distributed Model Support:} Larger models can be partitioned across multiple nodes (cloud, edge, RIC) to meet latency and compute constraints. The marketplace orchestrates these segments as a unified service. 
    
    \item \textbf{Collaborative Model Inference:}
    Multi-agent or mixture-of-experts GAI can be set up, where specialized “submodels” handle different request types. The platform coordinates routing and token-level streaming among these submodels, exploiting flexible compute availability in the network.

    \item \textbf{Service Mesh:} 
    A service mesh ensures efficient and secure operation of AI models in standalone, distributed, and collaborative environments. It manages communication between model components using proxies like Envoy, provides intelligent traffic management, enhances security with mutual transport layer security (TLS) encryption, and offers real-time observability of AI services. 
    
    Service mesh supports the complex interactions needed for collaborative AI models, ensuring that the marketplace can deliver scalable, high-performance, and secure AI services across telecoms networks.
    
    \item \textbf{UE-led Generation:} In use cases demanding near-instant feedback (AR/VR, gaming), the user equipment (UE) could drive the prompt generation while the marketplace provides localized GAI inference. The marketplace’s flexible billing captures usage patterns even at the UE level.
        
\end{itemize}

\subsection{Monetization Process}
Fig.~\ref{fig: marketplace monetization process} illustrates how the telecom GAI marketplace handles monetization from end to end. A GAI request (from a user device, xApp, or third-party application) enters the marketplace on the left. The GAI service/model provider pre-sets resource and policy constraints (e.g., compute, latency, QoS, and wireless resources), which define the subscription plan. The marketplace then checks whether the requestor’s plan is valid for those constraints. If the plan is insufficient, the request is rejected. Otherwise, it proceeds to the in-network GAI model from different routes for inference.

Once the model generates an output, the marketplace records usage metrics such as token count, GPU time, and data traffic. These metrics feed into the billing engine, which calculates costs accordingly. The Payment Gateway finalizes the transaction, and an optional revenue-sharing mechanism compensates the provider. By tying resource constraints directly to usage data, the marketplace enforces policy compliance and simplifies monetization for both operators and third-party GAI model providers.

\subsection{Implementation Framework}
\begin{figure}[t]
    \centering
    \includegraphics[width=1.0\linewidth]{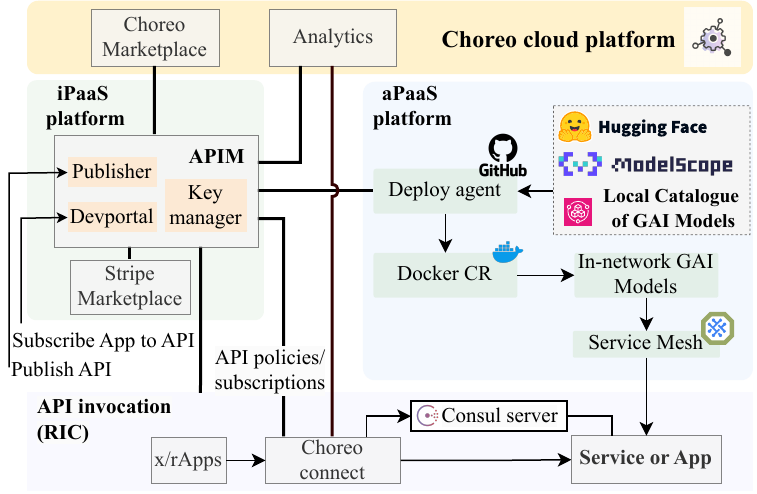}
    \caption{Overall marketplace integration architecture with API invocation. Multiple distributed or collaborative GAI models can be handled seamlessly via centralized API publishing, management, and usage-based billing.}
    \label{fig: marketplace_implementation}
\end{figure}

The marketplace implementation builds on the strategies outlined in~\cite{10355639}, originally designed for the Open RAN ecosystem. It emphasizes an API-centric integration Platform-as-a-Service (iPaaS) model to facilitate the seamless integration, deployment, and monetization of GAI models, applications (x/rApps), and other services within a 5G Open RAN infrastructure, leveraging Open RAN API standards. The key design features of this marketplace are as follows:
\begin{enumerate}
    \item API-Centric iPaaS Model: The marketplace uses an iPaaS approach, focusing on API-based integration to enable flexible deployment and monetization across different services and environments. This model provides fine-grained access control and monitoring, supporting various business models like pay-per-use, subscriptions, and SLAs.

    \item Multiple Deployment Environments: The marketplace supports various runtime environments--edge, cloud, and hybrid setups. Deployment agents automate service deployment across these environments, improving flexibility and optimization.

    \item Integration Features: It utilizes WSO2 API management to enable easy integration and deployment of applications. API gateways using Choreo Connect and Envoy proxies manage access and ensure secure service communication. It implements a federated service mesh to facilitate secure interactions between cloud and edge data centers.

    \item Monitoring and Reconciliation: It incorporates robust monitoring and billing mechanisms with tools like Stripe marketplace plugins, Choreo analytics, and Hyperledger blockchain for decentralized auditing, ensuring accurate performance tracking, billing, and compliance.
\end{enumerate}

Fig.~\ref{fig: marketplace_implementation} illustrates the overall marketplace integration architecture. The WSO2 API manager handles the marketplace management APIs and portals, enabling API publishing and subscription, and the creation of API product bundles that represent integrated applications utilizing multiple GAI services. These bundles are propagated to marketplace billing platforms Stripe, allowing for the provision of optimized, ready-to-use services for specific application use cases. This is particularly valuable for non-developer users who prefer not to handle selection, evaluation, testing, and service integration themselves.

GAI services and product bundles are deployed using YAML scripts from GitHub and containers from Docker Hub, Hugging Face, or other repositories. The Kubernetes-based YAML scripts used by the deployment agents are designed for low complexity and flexibility, featuring annotations for automated injection of service mesh sidecars and security configurations for different deployment environments. This approach allows users to integrate services without needing to be code developers.

The Federated Consul service mesh is used to securely manage service interactions, with Consul server instances deployed for each environment. This lightweight and highly scalable solution is available as an open-source addition.
For cross-tenant integration between isolated service mesh clusters deployed in different environments, the Choreo microgateway is used, leveraging the lightweight Envoy proxy.

This GAI marketplace is highly scalable, leveraging cloud-based service distribution to dynamically allocate resources across edge, near-edge, and central cloud environments.

\begin{figure}[t]
    \centering
    \includegraphics[width=1.0\linewidth]{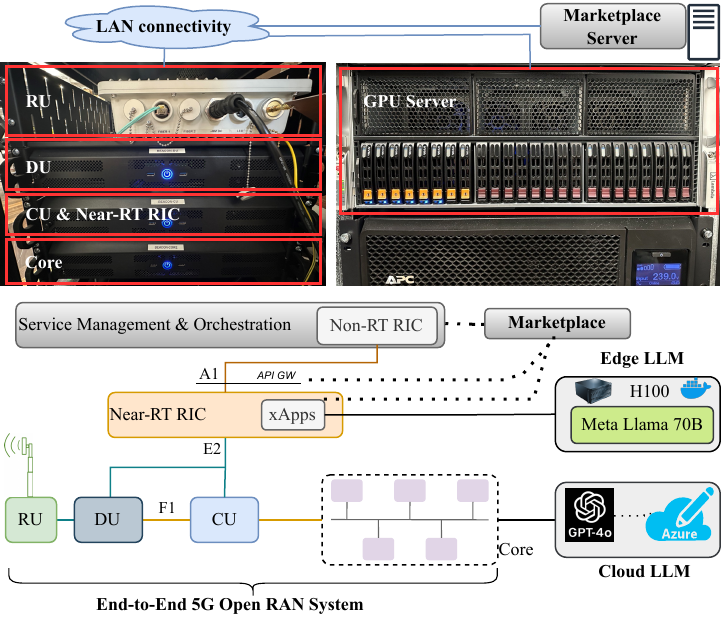}
    \caption{Illustration of the experimental setup. The local LLM container is deployed on the edge server near the Open RAN CU. Requests are managed via the marketplace's orchestrator.}
    \label{fig: experiment_overview}
\end{figure}
\section{Case Study: Local LLM Deployment in Open RAN testbed through Marketplace}

To illustrate the marketplace approach, we present a proof-of-concept scenario where a local large language model is deployed at the network edge, managed via the telecoms marketplace. We compare its performance to cloud-based LLM endpoints under realistic loading.

\subsection{Experimental Setup}
The experiment has been conducted using an advanced end-to-end Open RAN testbed developed under the BEACON-5G project~\cite{aijaz2023open}. The setup, illustrated in Fig.~\ref{fig: experiment_overview}, included an Open RAN system, and the LLM service agent which is responsible for handling requests and running GAI models. The LLM agent was deployed on an edge server linked to the Centralized Unit (CU) of the Open RAN. The management of the marketplace, including deployment and API billing components, was handled by the Non-Real-Time RIC (Non-RT RIC).
\begin{figure*}[t]   
    \subfloat[\label{fig:input_histogram}]{
      \begin{minipage}[t]{0.47\linewidth}
        \centering 
        \includegraphics[width=3.5in]{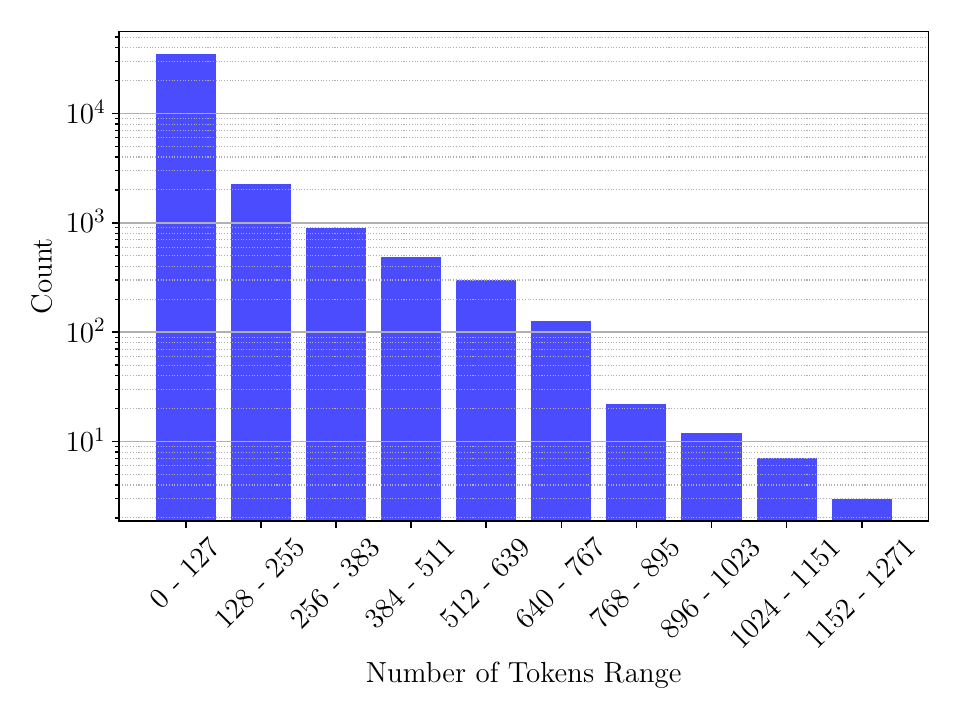}   
      \end{minipage}%
      }
      \hfill
        \subfloat[\label{fig:output_histogram}]{
      \begin{minipage}[t]{0.47\linewidth}   
        \centering   
        \includegraphics[width=3.5in]{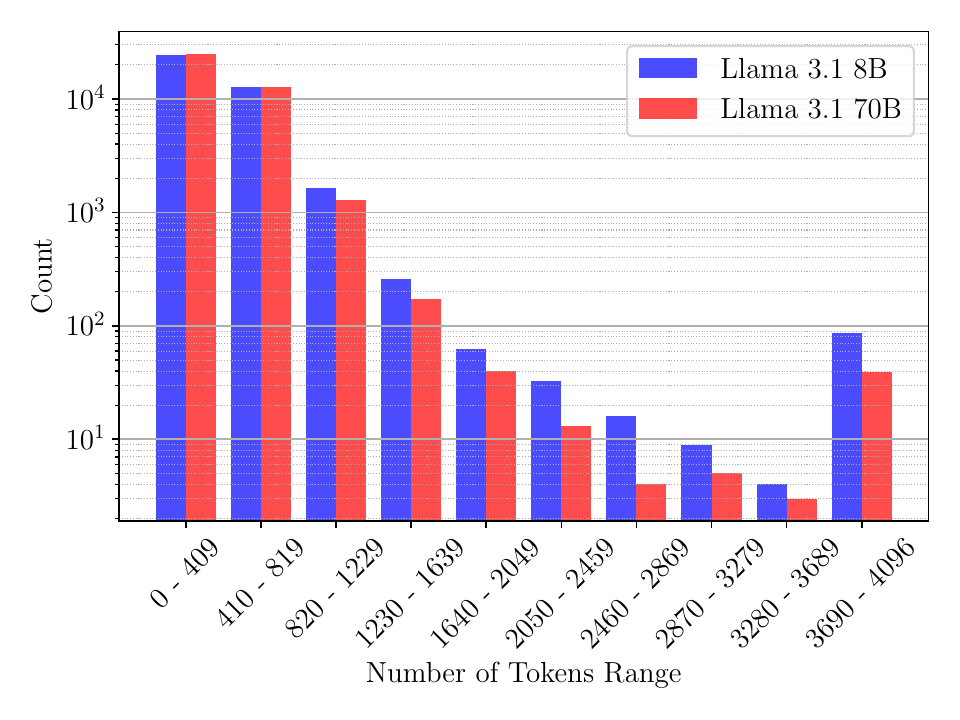}   
      \end{minipage} 
      }
      \caption{Histograms of Input and Output token lengths for the Chatbot arena dataset with the Llama 3.1 8B and 70B models, wherein (a) indicates Log-scaled Histogram of Input Tokens, (b) show Log-scaled Histogram of Output Tokens for Llama 3.1 Models.
      } \label{fig:token_histograms}
\vspace{-1.45mm}
\end{figure*} 

\subsubsection{Hardware and Software Specifications}
The edge server is equipped with two Intel Xeon Platinum 8480+ CPUs (shared access) and two Nvidia H100 PCIe GPUs, each with 80 GB VRAM. It hosts the vLLM agent~\cite{kwon2023efficientmemorymanagementlarge}. LLM Models are deployed in a vLLM container (one at a time) for high-performance inference compatible with OpenAI APIs:
    \textbf{Meta Llama 3.1 8B}~\cite{dubey2024llama3herdmodels} is used for fast, high-throughput processing of latency-sensitive tasks with moderate reasoning needs.
    \textbf{Meta Llama 3.1 70B} is deployed for medium-throughput processing of complex reasoning tasks to evaluate edge performance.

For the cloud deployment, the following models were deployed to Azure Cloud Cognitive Services:
    \textbf{GPT-3.5 Turbo} is deployed to an inference endpoint to serve as the cloud-based LLM for simple and quick tasks, equivalent in reasoning performance to the Llama 3.1 8B local model.
    \textbf{GPT4o} is deployed to an inference endpoint to serve as the cloud-based LLM for complex tasks, slightly superior in reasoning performance to the Llama 3.1 70B local model.

\subsubsection{Load Simulation}
To simulate a realistic inference environment, a constant background load was applied to the edge inference server using user-generated, chat-based open-response requests from the Chatbot Arena Dataset~\cite{zheng2023judging}. The system generated an average of 10 requests per second for the Llama 3.1 8B model and 3 requests per second for the Llama 3.1 70B model, with request intervals following an exponential distribution to simulate a Poisson process.

This setup resulted in an average of 44 concurrent background requests for the 8B model and 42 for the 70B model. The average input prompt size was 85 tokens, with the 8B model generating an average of 351 tokens per request and the 70B model generating 327 tokens on average.

As shown in Fig.~\ref{fig:input_histogram}, the distribution of input token lengths from the Chatbot Arena dataset is concentrated at lower token counts. In contrast, Fig.~\ref{fig:output_histogram} generally follows a similar trend but has a slightly higher concentration in 3690-4096 range due to the maximum token limit of 4096, which leads to output truncation and accumulation in the final bin. The 8B model also tends to generate longer responses, though this behavior was not specifically analyzed in this study.

\subsubsection{Request Configuration}
The LLM models in both the edge and cloud setups were configured to handle requests with specific parameters. The input tokens were set to 10, while the maximum output tokens allowed were 1000. Streaming functionality was enabled, and where applicable, the models were configured to ignore the end of sequence (EOS) token.

\subsection{Testing Procedure}
To compare cloud-based and edge-based LLM deployments in terms of latency during content generation tasks, identical requests were sent to both under controlled conditions, and key latency metrics were assessed.

\subsubsection{Measurement Metrics}

Time to First Token (TFT) measures the time from when a request is sent to the API server until the first token is received by the UE. This metric is crucial for applications requiring quick response times, such as interactive systems and real-time communications.

Inter-Token Time (ITT) measures the interval between successive tokens in the response stream. Consistent ITT is essential for applications that rely on continuous data streams, ensuring stable output rates~\footnote{We focus on local-deployment benefit metrics regarding latency and response times for PoC illustration. Other models, hardware configurations and inference engines would provide varying results.}.

\begin{figure}[t]
    \centering
    \includegraphics[width=1.0\linewidth]{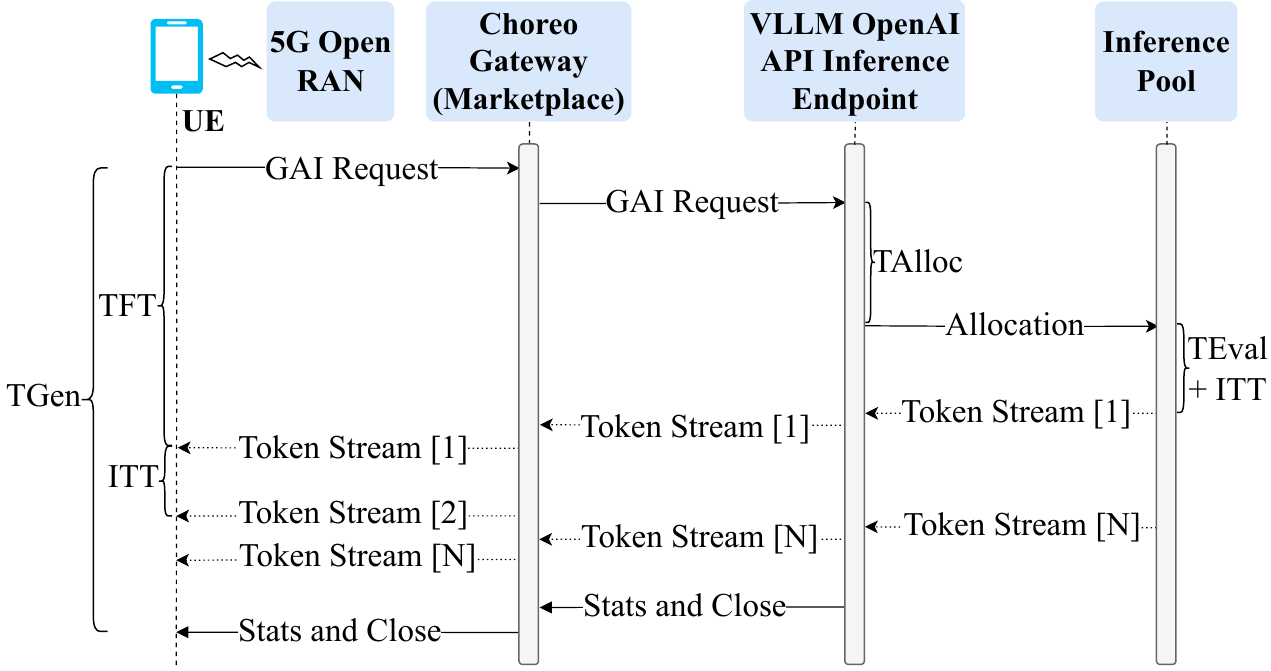}
    \caption{Illustration of the process of GAI content generation over marketplace platform and Open RAN testbed and time measurements, wherein TFT and ITT are the key metrics for comparison.}
    \label{fig:time_measurements}
\end{figure}

The GAI content generation process and timing measurements are illustrated in Fig.~\ref{fig:time_measurements}.

\subsubsection{Experimental Procedure}
The experiment was designed with several key considerations. First, 100 identical requests were sent sequentially to both edge-based and cloud-based LLMs. The responses were streamed back to the UE, and the TFT and ITT were recorded for each token generated. To simulate real-world conditions, a continuous background load was maintained on the edge server. The TFT and ITT metrics were averaged over multiple runs, and the variations were analyzed to assess the consistency and reliability of both deployment strategies.

\subsection{Results}

The comparison focused on cloud-based versus edge-based LLM deployments through the marketplace, highlighting two key scenarios:

\subsubsection{GPT-3.5 Turbo vs. Llama 3.1 8B}
As illustrated in Fig.~\ref{fig:results_gpt35_turbo_vs_llama8b}, the edge-based Llama 3.1 8B model outperformed the cloud-based GPT-3.5 Turbo on TFT whilst being slightly worse on ITT. The reason for this is the harsher synthetic load that we have set for the smaller 8B model assuming a higher usage. The Llama 3.1 8B model demonstrated lower TFT, providing faster initial responses, which is critical for time-sensitive applications like real-time interactive systems. Additionally, whilst the edge deployment exhibited inferior ITT, it had a reduced variability, making it well-suited for tasks requiring quick and reliable outputs, such as short text generation and classification. 

\subsubsection{GPT-4o vs. Llama 3.1 70B}
\begin{figure*}[t]   
    \subfloat[\label{fig:results_gpt35_turbo_vs_llama8b}]{
      \begin{minipage}[t]{0.47\linewidth}
        \centering 
        \includegraphics[width=3.5in]{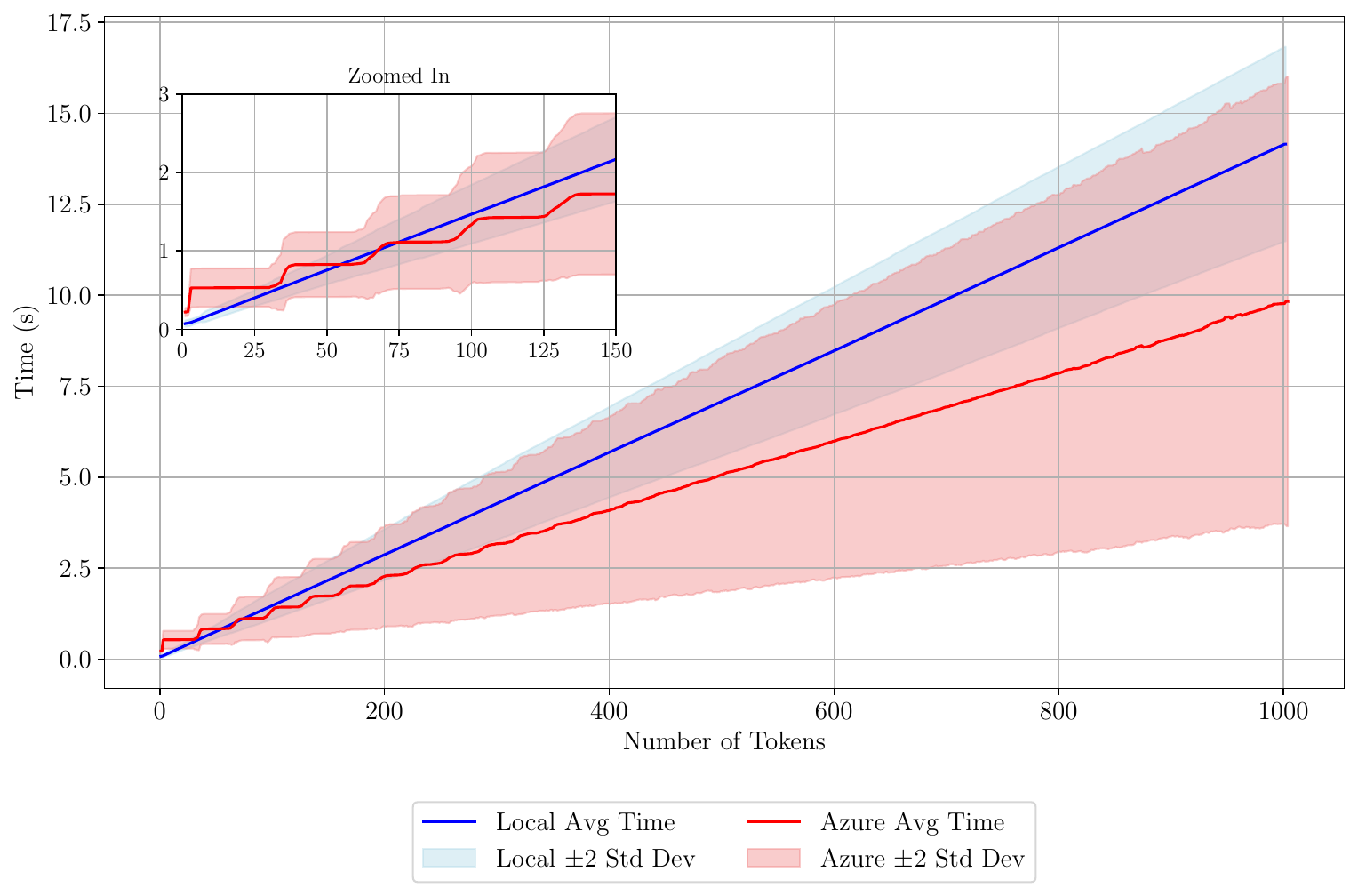}   
      \end{minipage}%
      }
      \hfill
        \subfloat[\label{fig:results_gpt4o_vs_llama70b}]{
      \begin{minipage}[t]{0.47\linewidth}   
        \centering   
        \includegraphics[width=3.5in]{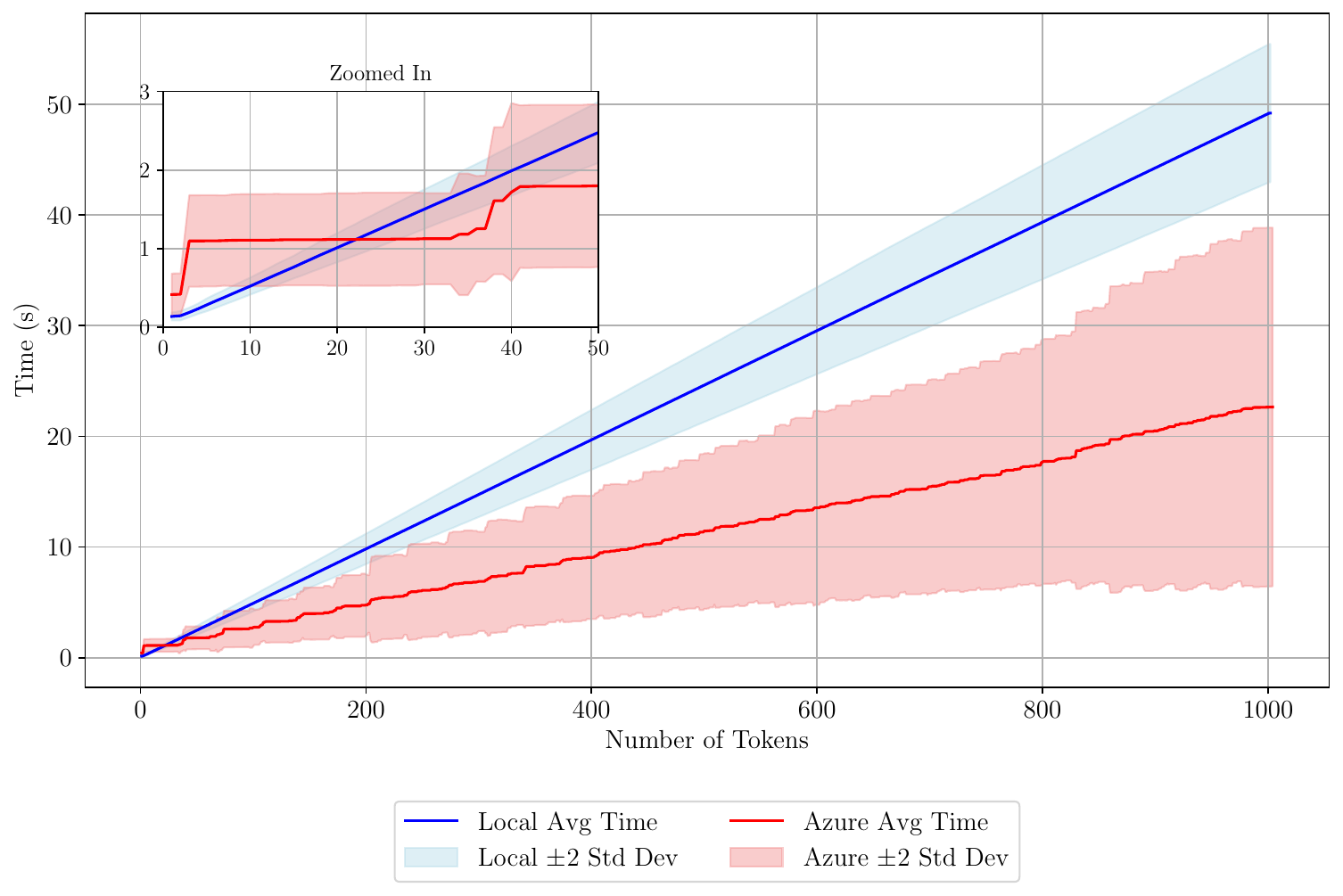}   
      \end{minipage} 
      }
      \caption{Token timing comparisons for different models and deployment environments. (a) shows token timing comparison between cloud-based GPT-3.5 Turbo and edge-based Llama 3.1 8B models; (b) shows token timing comparison between cloud-based GPT-4o and edge-based Llama 3.1 70B models.
      } \label{fig:combined_token_timing}
\vspace{-1.45mm}
\end{figure*}

Fig.~\ref{fig:results_gpt4o_vs_llama70b} presents a more nuanced comparison between the GPT-4o and Llama 3.1 70B models, showcasing the trade-offs between edge and cloud deployments. Similar to the smaller models, the edge-based Llama 3.1 70B exhibited a lower TFT, making it advantageous for quick, short responses, which is ideal for applications such as classification and short-form content generation. 

On the other hand, the cloud-based GPT-4o excelled in ITT, delivering faster and more consistent token generation for longer prompt completions. This makes it a better choice for tasks involving complex and lengthy content generation, where sustained output is more important than initial response speed. 

These findings underscore the importance of selecting the appropriate deployment strategy based on the specific requirements of the application, balancing the trade-offs between initial response speed and sustained generation performance. In this process, the GAI marketplace functions as a GAI service gateway, responsible for selecting, executing, and billing the appropriate GAI model. This process can be automated by the given application scenarios.

\section{Discussions}
\subsection{Monetization Strategy}

The marketplace should allow multiple charging models. Beyond conventional \emph{per-API call} or \emph{per-token} billing, it can incorporate:
\begin{itemize}[leftmargin=*]
    \item \textbf{Resource Accounting.} Operators track GPU hours, memory, or bandwidth used, enabling usage-based cost sharing among tenants. 
    \item \textbf{Vertical Bundles.} GAI-based analytics or chatbots can be combined with 5G/6G data plans or edge slices. This fosters synergy between network subscription revenues and AI services.
    \item \textbf{Revenue Sharing.} Third-party developers receive royalties if their adapters or models are actively invoked by paying clients. 
\end{itemize}
Token-level usage logs and policy enforcement can be implemented via the marketplace’s analytics, ensuring transparent cost reporting for all stakeholders.

\subsection{Service Access and Regulation}
The telecoms GAI marketplace presents a complex yet promising framework for integrating GAI services into telecom networks. It facilitates the deployment of AI solutions directly within the network infrastructure, enabling dynamic scaling and updates based on demand and performance. This flexibility allows MNOs to adapt quickly to changing market needs and customer expectations.

By centralizing model access in the marketplace, MNOs can more easily enforce data policies (e.g., restricting certain traffic to local edges for privacy) and authenticate/authorize third-party usage. However, operators must still address:
\begin{itemize}[leftmargin=*]
    \item \emph{Privacy compliance.} Potentially performing partial or fully anonymized inference. 
    \item \emph{Data residency.} Ensuring that GAI processing occurs within mandated geographic boundaries, especially for sensitive enterprise or government customers.
    \item \emph{Fairness and transparency.} Auditing generative outputs for bias or harmful content. 
\end{itemize}

\subsection{Service Management} 
Managing the lifecycle of GAI services in the marketplace requires careful planning and robust strategies. This includes monitoring performance, handling updates and patches, ensuring service continuity, and optimizing resource allocation. Effective management is essential for maintaining reliable and efficient GAI services, which directly affects customer satisfaction and operational costs. For MNOs, balancing these technical requirements with strategic goals like cost management, service innovation, and market competitiveness is vital to fully leveraging the telecoms GAI marketplace.

\subsection{Open Service Platform for Marketplace}
Open networks require the collaboration of multiple operators and service providers. In this context, the open service platform (OSP) is an option for coordinating network operations and GAI services. Integrating the OSP with the private marketplace provides a unified access point for service provisioning, enabling dynamic and flexible service offerings that can quickly respond to customer demands and operational needs.

Aligning the OSP with marketplace mechanisms allows customers to easily discover, purchase, and configure services across different networks. The OSP also plays a key role in ensuring fairness by balancing GAI services from various providers, enabling comparisons, and detecting anti-competitive pricing strategies. This functions like a price comparison tool but with enhanced monitoring and evaluation of integrated services.

The combination of the marketplace and OSP can enhance collaboration among service providers. This allows operators to expand their service offerings without significant infrastructure investments, enabling scalable and flexible deployment of GAI services across diverse network environments.

\section{Conclusions and future directions}
In this article, we have presented a telecoms GAI marketplace solution aimed at maximizing the benefits of large generative AI models in open 6G networks. From an MNO standpoint, our approach addresses key hurdles in \emph{monetization, management,} and \emph{service quality}. By hosting, orchestrating, and billing GAI services directly within the network, operators can reduce dependencies on external cloud providers, gain finer control over latency and data governance, and co-create novel service bundles with third-party developers.

Our proof-of-concept deployment in an Open RAN testbed shows promising latency advantages for local LLMs. Cloud-based solutions may retain strengths in sustained token throughput, underscoring the need for an intelligent marketplace that routes requests according to specific application demands. The marketplace paradigm is readily extensible to future multi-modal models and advanced inference pipelines.

Future work can delve into collaborative AI approaches (e.g., mixture-of-experts or hierarchical inference across multiple edge servers), multi-operator interoperability, and advanced data privacy frameworks such as homomorphic encryption or differential privacy for GAI tasks. Close alignment with evolving 6G standards and open network specifications (e.g., O-RAN Alliance, 3GPP, AI-RAN Alliance) will be critical to ensure that telecoms GAI marketplaces seamlessly integrate into next-generation architectures. We hope this article lays a foundation for broader adoption of generative AI in open 6G networks, ultimately driving both technical innovation and new revenue opportunities for MNOs.

\bibliographystyle{IEEEtran} %

\bibliography{IEEEabrv,references} 

\end{document}